\documentclass[aps,prl,twocolumn,showpacs,floatfix,superscriptaddress]{revtex4-1}
\usepackage{graphicx,amsfonts,amssymb,amsmath,hyperref}
\usepackage{textcomp}
\usepackage{esint}

\newif\ifhyper
\hypertrue
\ifhyper       
\hypersetup{        
  citecolor = {green},
  colorlinks = {true}, % false
  urlcolor = {blue} % magenta
}

\begin{document}

%AUTHOR'S MACRO
\def\Tkt{T_{\rm BKT}}

%%%%%%%%%%%%%%%%%%%%%%%%%%%%%%%%%%%%%%%%%%%

\graphicspath{{./figures/}}

\def\rhoeq{\hat\rho_{\rm eq}}

\newcommand{\marge}[1]{\marginpar{\scriptsize #1}}
\newcommand{\remarque}[1]{\marginpar{\scriptsize Remarque}{\it [#1]}}
\newcommand{\red}[1]{\textcolor{red}{#1}}
\newcommand{\blue}[1]{\textcolor{blue}{#1}}

\def\beq{\begin{equation}}
\def\eeq{\end{equation}}
\def\bleq{\begin{eqnarray}}
\def\eleq{\end{eqnarray}} 
\def\bfig{\begin{figure}}
\def\efig{\end{figure}}
\def\bline{\begin{multline}}
\def\eline{\end{multline}}
\def\bremark{\begin{quotation} \noindent \small }
\def\eremark{\end{quotation}}
\def\llbrace{\left\lbrace}
\def\rrbrace{\right\rbrace}
\def\rrangle{\right\rangle}
\def\llangle{\left\langle}
\def\lbraket{\left[}
\def\rbraket{\right]}

\newcommand{\Tr}{{\rm Tr}} 
\newcommand{\tr}{{\rm tr}} 
\newcommand{\sgn}{{\rm sgn}} 
\newcommand{\mean}[1]{\langle #1 \rangle}
\newcommand{\commu}[2]{[#1,#2]} 
\newcommand{\bra}[1]{\langle#1|}
\newcommand{\ket}[1]{|#1\rangle}
\newcommand{\braket}[2]{\langle #1|#2\rangle}
\newcommand{\dbraket}[3]{\langle #1|#2|#3\rangle}
\newcommand{\tens}[1]{\overleftrightarrow{#1}}  
\newcommand{\vac}{|{\rm vac}\rangle} 
\def\bravac{\langle{\rm vac}|}
\newcommand{\const}{{\rm const}} 
\newcommand{\atanh}{\,{\rm atanh}}

\newcommand{\ie}{i.e. }
\newcommand{\iet}{i.e.}
\newcommand{\eg}{e.g. }
\newcommand{\cc}{{\rm c.c.}} 
\newcommand{\hc}{{\rm h.c.}} 
\def\etal{{\it et al. }}

\newcommand{\jhatbf}{\hat {\textbf \j}} 
\newcommand{\Jhatbf}{\hat {\textbf \J}} 
\newcommand{\jhat}{\hat {\jmath}} 
\newcommand{\Jhat}{\hat {J}} 
\newcommand{\jbf}{\textbf j}
\newcommand{\Jbf}{\textbf J}

\def\chibf{\boldsymbol{\chi}}
\def\down{\downarrow}
\def\eps{\epsilon}
\def\gam{\gamma} 
\def\phibf{\boldsymbol{\phi}}
\def\varphibf{\boldsymbol{\varphi}}
\def\varphibfs{\boldsymbol{\varphi}_<}
\def\varphibfl{\boldsymbol{\varphi}_>}
\def\varphis{\varphi_{<}}
\def\varphil{\varphi_{>}}
\def\psibf{\boldsymbol{\psi}}
\def\Ome{\Omega}
\def\omeD{{\omega_D}} 
\def\bfOme{\boldsymbol{\Omega}} 
\def\Omebf{\boldsymbol{\Omega}} 
\def\lamb{\lambda}
\def\Lamb{\Lambda}
\def\sig{\sigma}
\def\Sig{\Sigma}
\def\sigp{{\sigma'}} 
\def\bfsig{\boldsymbol{\sigma}} 
\def\sigbf{\boldsymbol{\sigma}} 
\def\The{\Theta} 
\def\up{\uparrow}

\def\epsk{\epsilon_{\bf k}} 
\def\xik{\xi_{\bf k}} 
\def\xip{\xi_{\bf p}}
\def\xikq{\xi_{{\bf k}+{\bf q}}} 
\def\Ek{E_{\bf k}}
\def\Ep{E_{\bf p}}
\def\Heff{\hat H_{\rm eff}}
\def\Hem{\hat H_{\rm em}}
\def\Hint{\hat H_{\rm int}}
\def\Hloc{\hat H_{\rm loc}}
\def\HMF{\hat H_{\rm MF}}
\def\Sem{S_{\rm em}}
\def\SMF{S_{\rm MF}} 
\def\SRPA{S_{\rm RPA}} 
\def\Sint{S_{\rm int}} 
\def\Sloc{S_{\rm loc}} 
\def\Zloc{Z_{\rm loc}} 
\def\ZMF{Z_{\rm MF}} 
\def\ZRPA{Z_{\rm RPA}} 
\def\RPA{{\rm RPA}}
\def\loc{{\rm loc}} 
\def\pp{{\rm pp}}
\def\ph{{\rm ph}} 
\def\ch{{\rm ch}}
\def\sp{{\rm sp}} 
\def\qtf{q_{\rm TF}}
\def\epstf{\eps^{}_{\rm TF}} 
\def\epsrpa{\eps^{}_{\rm RPA}} 
\def\chinnzpp{\chi_{nn}^{0}{}\!\!\!''}

\def\half{\frac{1}{2}}
\def\dhalf{\dfrac{1}{2}}
\def\third{\frac{1}{3}} 
\def\quarter{\frac{1}{4}}

\def\qr{{\bf q}\cdot{\bf r}}
\def\wt{\omega t} 

\def\a{{\bf a}}
\def\b{{\bf b}}
\def\e{{\bf e}}
\def\f{{\bf f}}
\def\g{{\bf g}}
\def\h{{\bf h}}
\def\k{{\bf k}}
\def\l{{\bf l}}
\def\m{{\bf m}}
\def\n{{\bf n}} 
\def\p{{\bf p}} 
\def\q{{\bf q}}
\def\r{{\bf r}}
\def\t{{\bf t}}
\def\u{{\bf u}}
\def\v{{\bf v}}
\def\x{{\bf x}}
\def\y{{\bf y}} 
\def\z{{\bf z}} 
\def\A{{\bf A}}
\def\B{{\bf B}}
\def\D{{\bf D}} 
\def\E{{\bf E}} 
\def\F{{\bf F}} 
\def\H{{\bf H}}  
\def\J{{\bf J}}
\def\K{{\bf K}} 

\def\G{{\bf G}}
\def\L{{\bf L}}
\def\M{{\bf M}}  
\def\O{{\bf O}} 
\def\P{{\bf P}} 
\def\Q{{\bf Q}} 
\def\R{{\bf R}}
\def\S{{\bf S}}
\def\epsbf{\boldsymbol{\epsilon}}
\def\mubf{\boldsymbol{\mu}}
\def\nablabf{\boldsymbol{\nabla}}
\def\rhobf{\boldsymbol{\rho}}
\def\sigmabf{\boldsymbol{\sigma}} 
\def\Pibf{\boldsymbol{\Pi}}
\def\pibf{\boldsymbol{\pi}}

\def\para{\parallel}
\def\kpara{{k_\parallel}}
\def\kperp{{k_\perp}} 
\def\kperpp{{k_\perp'}} 
\def\qperp{{q_\perp}} 
\def\tperp{{t_\perp}} 

\def\w{\omega}
\def\wn{\omega_n}
\def\wnu{\omega_\nu}
\def\wp{\omega_p} 
\def\dmu{{\partial_\mu}}
\def\dl{{\partial_l}}  
\def\dt{\partial_t} 
\def\tdt{\tilde\partial_t}
\def\dk{\partial_k}
\def\tdk{\tilde\partial_k}
\def\dx{\partial_x}
\def\dy{\partial_y} 
\def\dtau{{\partial_\tau}}  
\def\det{{\rm det}} 
\def\Pf{{\rm Pf}}

\def\dsum{\displaystyle \sum}
\def\dint{\displaystyle \int} 
\def\intt{\int_{-\infty}^\infty dt} 
\def\inttp{\int_{-\infty}^\infty dt'} 
\def\intk{\int_{\bf k}} 
\def\intkd{\int \frac{d^dk}{(2\pi)^d}}
\def\intq{\int_{\bf q}} 
\def\intr{\int d^dr}  
\def\dintr{\displaystyle \int d^dr} 
\def\intrp{\int d^dr'}
\def\dinttau{\displaystyle \int_0^\beta d\tau}
\def\dinttaup{\displaystyle \int_0^\beta d\tau'}
\def\inttau{\int_0^\beta d\tau}
\def\inttaup{\int_0^\beta d\tau'}
\def\intx{\int d^{d+1}x} 
\def\inttaur{\int_0^\beta d\tau \int d^dr}
\def\intinf{\int_{-\infty}^\infty}
\def\dinttaur{\displaystyle \int_0^\beta d\tau \int d^dr}
\def\dintinf{\displaystyle \int_{-\infty}^\infty}
\def\intw{\int_{-\infty}^\infty \frac{d\w}{2\pi}}
\def\sumr{\sum_{\bf r}} 

\def\calA{{\cal A}} 
\def\calC{{\cal C}} 
\def\dt{\partial_t}
\def\calD{{\cal D}}
\def\calF{{\cal F}} 
\def\calG{{\cal G}}
\def\calH{{\cal H}}
\def\calI{{\cal I}}
\def\calJ{{\cal J}}
\def\calK{{\cal K}}
\def\calL{{\cal L}} 
\def\calN{{\cal N}}
\def\calO{{\cal O}}
\def\calP{{\cal P}}  
\def\calR{{\cal R}} 
\def\calS{{\cal S}}
\def\calT{{\cal T}}
\def\calU{{\cal U}}
\def\calX{{\cal X}}
\def\calY{{\cal Y}} 
\def\calZ{{\cal Z}} 

\def\calFbf{{\bf F}}

\def\tT{{\tilde T}}
\def\talpha{{\tilde\alpha}}
\def\tdelta{{\tilde\delta}}
\def\teta{{\tilde\eta}} 
\def\tlamb{{\tilde\lambda}}
\def\tmu{{\tilde\mu}}
\def\tphibf{{\tilde\phibf}}
\def\trho{{\tilde\rho}}
\def\tvarphibf{{\tilde\varphibf}} 
\def\tw{{\tilde\omega}}
\def\twn{{\tilde\omega_n}}

\def\asinh{{\rm asinh}} 

\title{Higgs amplitude mode in the vicinity of a $(2+1)$-dimensional quantum critical point}  

\author{A. Ran\c{c}on}
\affiliation{James Franck Institute and Department of Physics,
University of Chicago, Chicago, Illinois 60637, USA}

\author{N. Dupuis}
\affiliation{Laboratoire de Physique Th\'eorique de la Mati\`ere Condens\'ee, 
CNRS UMR 7600, Universit\'e Pierre et Marie Curie, 4 Place Jussieu, 
75252 Paris Cedex 05, France}

\date{March 25, 2014} 

\begin{abstract}
We study the ``Higgs'' amplitude mode in the relativistic quantum O($N$) model in two space dimensions. Using the nonperturbative renormalization group we compute the O($N$)-invariant scalar susceptibility in the vicinity of the zero-temperature quantum critical point. In the zero-temperature ordered phase, we find a well defined Higgs resonance for $N=2$ with universal properties in agreement with quantum Monte Carlo simulations. The resonance persists at finite temperature below the Berezinskii-Kosterlitz-Thouless transition temperature. In the zero-temperature disordered phase, we find a maximum in the spectral function which is however not related to a putative Higgs resonance. Furthermore we show that the resonance is strongly suppressed for $N\geq 3$.
\end{abstract}
\maketitle 

At low temperatures many condensed-matter systems are described by a relativistic effective field theory with O($N$) symmetry ($N\geq 2$): quantum antiferromagnets, superconductors, Bose-Einstein condensates in optical lattices, etc. Far from criticality, collective excitations in these systems are in general well understood. In the disordered (symmetric) phase there are $N$ gapped modes. In the ordered phase, where the O($N$) symmetry is spontaneously broken, there are $N-1$ gapless Goldstone modes corresponding to fluctuations of the direction of the $N$-component quantum field, and a gapped amplitude (``Higgs'') mode~\cite{Sachdev_book}. 

The fate of the Higgs mode in low-dimensional systems near a (zero-temperature) quantum critical point (QCP) has been a subject of debate. Does the Higgs mode exist as a resonance-like feature or is it overdamped due to its coupling to Goldstone modes? In three dimensions, where the effective field theory is four-dimensional and noninteracting at the QCP, the Higgs resonance becomes sharper and sharper as the QCP is approached. This has been beautifully confirmed in the quantum antiferromagnet TlCuCl$_3$~\cite{Ruegg08} (see also Ref.~\cite{Bissbort11} for an experiment with cold atoms). In two space dimensions, the effective field theory is strongly coupled at the QCP and the existence of the Higgs resonance is not guaranteed. Furthermore the visibility of the Higgs mode strongly depends on the symmetry of the probe~\cite{Podolsky11}. The longitudinal susceptibility is dominated by the Goldstone modes and diverges as $1/\w$ at low frequencies~\cite{Sachdev99,Zwerger04,Dupuis09b}, 
thus making the observation of the Higgs resonance impossible. The O($N$)-invariant scalar susceptibility (i.e. the correlation function of the square of the order parameter field) has a spectral weight which vanishes as $\w^3$ and is a much better candidate~\cite{Podolsky11}. Quantum Monte Carlo (QMC) simulations of $(2+1)$-dimensional systems have shown that the Higgs resonance shows up in the scalar susceptibility and remains a well defined excitation arbitrarily close to the QCP for $N=2$ and $N=3$~\cite{Gazit13,Gazit13a,Pollet12,Chen13}. A signature of the Higgs mode has recently been observed in a two-dimensional Bose gas in an optical lattice in the vicinity of the superfluid--Mott-insulator transition~\cite{Endres12}.

In this Letter, we use a nonperturbative renormalization-group (NPRG) approach to the relativistic quantum O($N$) model and compute the scalar susceptibility near the QCP. We obtain the spectral function for arbitrary values of $N$. For $N=2$, we find a well defined Higgs resonance in the zero-temperature ordered phase with universal properties in good agreement with QMC simulations of a related model~\cite{Gazit13,Gazit13a} and the Bose-Hubbard model ~\cite{Pollet12,Chen13}. The resonance persists at finite temperature below the Berezinskii-Kosterlitz-Thouless (BKT) transition temperature; finite-temperature effects modify the spectral function only at low frequencies $\w\lesssim T$. Although we also find a maximum in the spectral function in the zero-temperature disordered phase, our RG analysis shows that this maximum cannot be interpreted as a Higgs resonance as recently suggested~\cite{Chen13}. Furthermore, we find that the resonance is strongly suppressed for $N\geq 3$. 

{\it Methods.} We consider the relativistic quantum O($N$) model defined by the (Euclidean) action 
\begin{align}
S[\varphibf,h] ={}&  \int dx \biggl\lbrace \half (\nablabf\varphibf)^2 + \frac{1}{2c^2} (\dtau \varphibf)^2 \nonumber \\ & 
+ \frac{r_0}{2} \varphibf^2 + \frac{u_0}{4!} {(\varphibf^2)}^2 - h \varphibf^2 \biggr\rbrace , 
\label{action} 
\end{align}
where we use the notations $x=(\r,\tau)$ and $\int dx=\inttau \int d^2r$. The $N$-component real field $\varphibf$ is periodic in the imaginary time $\tau$: $\varphibf(\r,\tau+\beta)=\varphibf(\r,\tau)$ ($\beta=1/T$ and we set $\hbar=k_B=1$). $r_0$ and $u_0$ are temperature-independent coupling constants and $c$ is the (bare) velocity of the $\varphibf$ field. We have added a uniform time-dependent source $h(\tau)$ which couples to $\varphibf(x)^2$ (and will eventually be set to zero). The model is regularized by an ultraviolet cutoff $\Lambda$ acting both on momenta and frequencies. 

\begin{figure}
\centerline{\includegraphics[width=4.75cm]{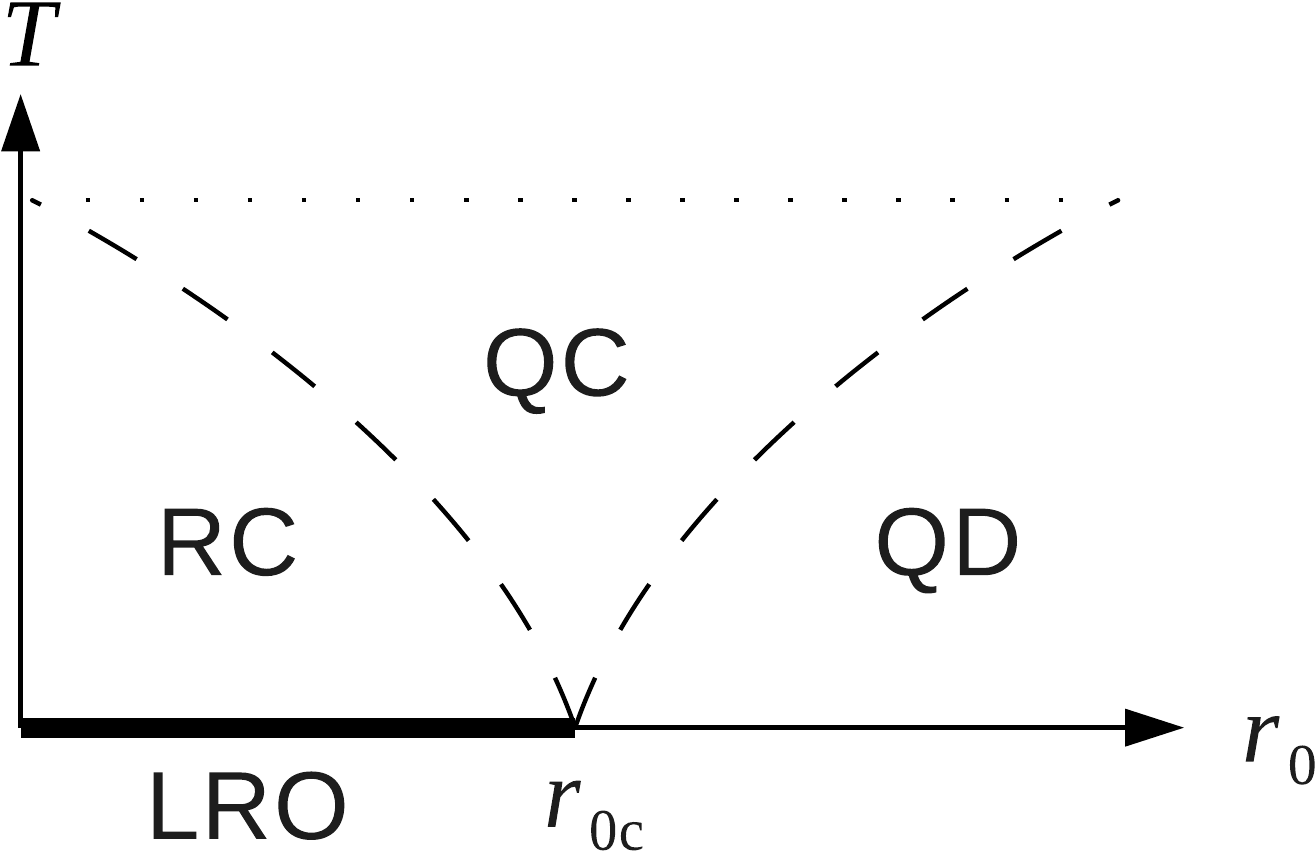}}
\caption{Phase diagram of the relativistic quantum O($N$) model in two space dimensions ($N\geq 2$). The thick line shows the zero-temperature ordered phase with long-range order (LRO), while the dashed lines are crossover lines between the renormalized classical (RC), quantum critical (QC) and quantum disordered (QD) regimes. The dotted line shows the limit of the high-$T$ region where the physics is not controlled by the QCP anymore. (For $N=2$, there is a finite-temperature BKT transition line for $r_0\leq r_{0c}$, which terminates at $T=0$ for $r_0=r_{0c}$.)}
\label{fig_phase_dia}
\end{figure}

In two space dimensions, the phase diagram of the relativistic quantum O($N$) model is well known (Fig.~\ref{fig_phase_dia})~\cite{Sachdev_book}. At zero temperature (and for $h=0$), there is a quantum phase transition between a disordered phase ($r_0>r_{0c}$) and an ordered phase ($r_0<r_{0c}$) where the O($N$) symmetry of the action~(\ref{action}) is spontaneously broken ($u_0$ and $c$ are considered as fixed parameters). The QCP at $r_0=r_{0c}$ is in the universality class of the three-dimensional classical O($N$) model with a dynamical critical exponent $z=1$ (this value follows from Lorentz invariance at zero temperature); the phase transition is governed by the three-dimensional Wilson-Fisher fixed point. At finite temperatures, the system is always disordered, in agreement with the Mermin-Wagner theorem, but it is possible to 
distinguish three regimes in the vicinity of the QCP depending on the temperature dependence of the correlation length $\xi(T)$: a renormalized classical regime where $\xi(T)\sim (c/\rho_s)e^{2\pi\rho_s/(N-2)T}$ (with $\rho_s$ the zero-temperature ``stiffness''), a quantum critical regime where $\xi(T)\sim c/T$, and a quantum disordered regime where $\xi(T)\sim\xi(0)$~\cite{Chakravarty89,Sachdev_book}. For $N=2$ and $r_0 < r_{0c}$, there is a finite-temperature Berezinskii-Kosterlitz-Thouless (BKT) phase transition~\cite{Berezinskii70,*Berezinskii71,*Kosterlitz73,*Kosterlitz74} and the system exhibits algebraic order at low temperatures. The BKT transition temperature line $\Tkt$ terminates at the QCP $r_0=r_{0c}$. 

The zero-momentum scalar susceptibility is defined by 
\begin{align}
\chi(i\wn) &= \int dx\, e^{i\wn\tau} \,\bigl[ \mean{\varphibf(x)^2 \varphibf(0)^2} - \mean{\varphibf(0)^2}^2 \bigr] \nonumber \\ 
&= \frac{\delta^2 \ln Z[h]}{\delta h(-i\wn) \delta h(i\wn)} \biggl|_{h=0} ,
\label{chi} 
\end{align}
where $Z[h]$ is the partition function obtained from the action~(\ref{action}) and $h(i\wn)$ the Fourier transform of $h(\tau)$. The spectral function 
\beq
\chi''(\w) = \mbox{Im}[\chi(i\wn\to \w+i0^+)] 
\eeq
is obtained by analytical continuation from Matsubara frequencies $\wn=2n\pi T$ ($n$ integer) to real frequencies $\w$. At zero temperature and in the universal regime near the QCP (scaling limit)~\cite{not4}, $\chi''$ takes the form~\cite{Podolsky12}
\beq
\chi''(\w) = \calA_\pm \Delta^{3-2/\nu} \Phi_\pm \left( \frac{\w}{\Delta} \right) , 
\label{scaling}
\eeq
where the index $+/-$ refers to the disordered and ordered phases, respectively. In the disordered phase, $\Delta\propto (r_0-r_{0c})^\nu$ is the gap in the excitation spectrum (with $\nu$ the correlation-length exponent at the QCP). In the ordered phase, $\Delta$ is defined as the gap at the mirror point (with respect to the QCP) in the disordered phase; the ratio $\Delta/\rho_s$ between $\Delta$ and the stiffness $\rho_s$ is an $N$-dependent universal number~\cite{Rancon13a}. $\calA_\pm$ is a nonuniversal cutoff-dependent constant while $\Phi_\pm(x)$ is a universal scaling function. For $\w\gg\Delta$, $\chi''(\w)\sim \w^{3-2/\nu}$ is independent of $\Delta$. In the ordered phase, the low-energy behavior $\w\ll\Delta$ is entirely determined by the Goldstone modes and $\chi''(\w)\sim \w^3$~\cite{Podolsky12}. In the disordered phase, the system is gapped and $\chi''(\w)$ vanishes for $\w<2\Delta $. (Since $\chi''(\w)=-\chi''(-\w)$ is odd, we discuss only the positive frequency part.) 

To implement the NPRG approach, we add to the action~(\ref{action}) an infrared regulator term $\Delta S_k[\varphibf]$ indexed by a momentum scale $k$ such that fluctuations are smoothly taken into account as $k$ is lowered from the microscopic scale $\Lambda$ down to zero~\cite{Berges02,Delamotte12,Kopietz_book}. This allows us to introduce the scale-dependent effective action 
\beq
\Gamma_k[\phibf,h] = - \ln Z_k[\J,h] + \int dx\, \J\cdot \phibf - \Delta S_k[\phibf] ,
\eeq
defined as a modified Legendre transform of the free energy $-\ln Z_k[\J,h]$ that includes the subtraction of $\Delta S_k[\phibf]$. Here $\phibf(x)=\delta \ln Z_k[\J,h]/\delta \J(x)=\mean{\varphibf(x)}$ is the order parameter and $\J$ an external source which couples linearly to the $\varphibf$ field. The variation of the effective action with $k$ is given by Wetterich's equation~\cite{Wetterich93} 
\beq
\dk \Gamma_k[\phibf,h] = \half \Tr\llbrace \dk R_k\left(\Gamma^{(2)}_k[\phibf,h] + R_k\right)^{-1} \rrbrace ,
\label{rgeq}
\eeq
where $\Gamma^{(2)}_k[\phibf,h]$ denotes the second-order functional derivative of $\Gamma_k[\phibf,h]$ with respect to $\phibf$. In Fourier space, the trace involves a sum over momenta and Matsubara frequencies as well as the O($N$) index of the $\phibf$ field. $R_k$ is a momentum-frequency dependent cutoff function appearing in the definition of the regulator term $\Delta S_k[\varphibf]$. The scalar susceptibility $\chi_k(i\wn)$ [Eq.~(\ref{chi})] is obtained from the second-order functional derivative of $\Gamma_k[\phibf,h]$ with respect to the source $h$~\cite{not1}.  

We solve the flow equation~(\ref{rgeq}) using two main approximation. The $h=0$ part of the effective action $\Gamma_k[\phibf,h]$ is solved using the Blaizot-Mendez-Wschebor approximation~\cite{Blaizot06,Benitez09,Benitez12} combined with a derivative expansion~\cite{Berges02,Delamotte12}. As for the $h$ dependent part, we use the following truncation,
\begin{multline}
\Gamma_k[\phibf,h] = \Gamma_k[\phibf,0] +  \half \int dy dy'\, h(y) H_k^{(0,2)}(y-y') h(y') \\
+ \int dx dy [\rho(x)-\rho_{0,k}] H^{(2,1)}_k(x-y) h(y) ,
\label{ansatz}
\end{multline} 
where we have introduced the O($N$) invariant $\rho=\phibf^2/2$ and its value $\rho_{0,k}$ at the minimum of $\Gamma_k[\phibf,h=0]$. We refer to the Supplemental Material for more details on the NPRG calculation of the scalar susceptibility~\cite{not1}. 

By numerically solving the flow equation for a given set of microscopic parameters ($\Lambda$, $u_0$, $r_0$, etc.), we obtain the scalar susceptibility $\chi(i\wn)\equiv \chi_{k=0}(i\wn)$. In practice, we compute $\chi(i\wn)$ for typically 50 or 100 frequency points and then use a Pad\'e approximant to deduce the spectral function $\chi''(\w)$~\cite{Vidberg77}. (For previous implementations of this method to compute dynamical correlations functions see Refs.~\cite{Dupuis09b,Dupuis09a,Sinner09,Sinner10}.)

\begin{figure}
\centerline{\includegraphics[width=6cm]{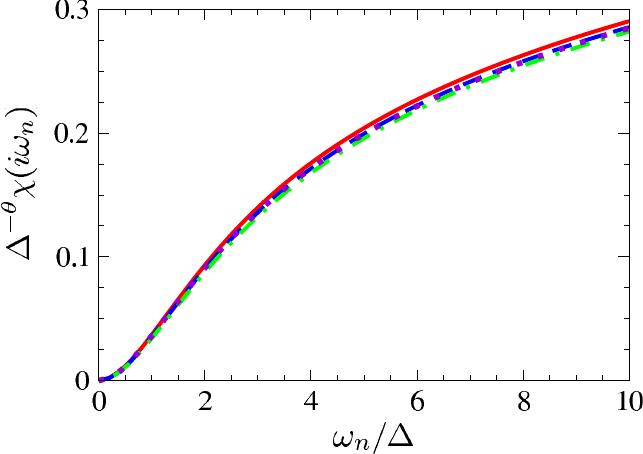}}
\caption{(Color online) $\Delta^{-\theta}[\chi(i\wn)-\chi(0)]$ vs $\wn/\Delta$ in the $T=0$ ordered phase ($N=2$) for  various values of $r_0$ ($0.001\leq r_{0c}-r_0\leq 0.005$ and $\Lamb=1$).} 
\label{fig_chi_matsubara}
\end{figure}

{\it Higgs resonance in the ordered phase.} Figure~\ref{fig_chi_matsubara} shows the scalar susceptibility $\chi(i\wn)$ in the $T=0$ ordered phase for $N=2$ and various values of $r_0$. By plotting $\Delta^{-\theta}[\chi(i\wn)-\chi(0)]$ as a function of the rescaled Matsubara frequency $\wn/\Delta$, we observe a data collapse in agreement with the expected universality. The subtraction of $\chi(0)$ eliminates a nonsingular nonuniversal constant~\cite{Gazit13}. We expect the exponent $\theta$ to be equal to $3-2/\nu\simeq 0.02248$ (using $\nu\simeq 0.6717$ for the three-dimensional O(2) model); within our approximations we find $\theta\simeq 0.1361$. Higher-order truncations of the effective action $\Gamma_k[\phibf,h]$ change the value of $\theta$ but hardly the shape of the Higgs resonance in the spectral function. 

\begin{figure}
\centerline{\includegraphics[height=4cm]{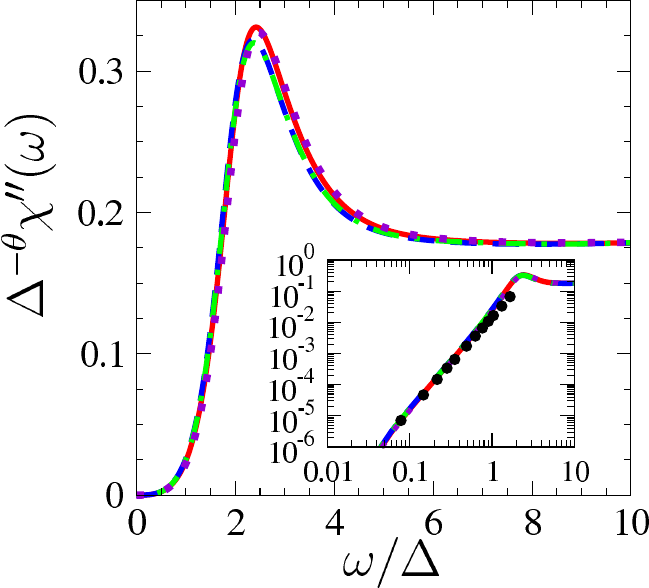}
\includegraphics[height=4cm]{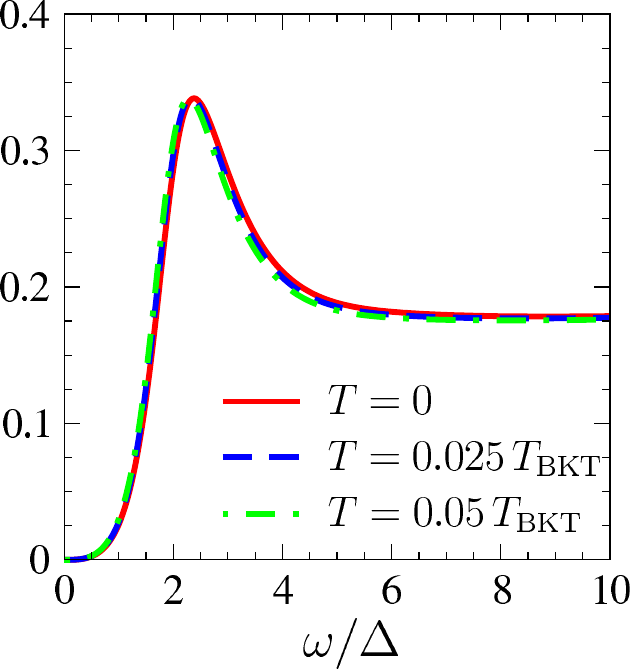}}
\caption{(Color online) Left: spectral function $\Delta^{-\theta}\chi''(\w)$ vs $\w/\Delta$ in the $T=0$ ordered phase ($N=2$) for various values of $r_0$ ($0.001\leq r_{0c}-r_0\leq 0.005$ and $\Lamb=1$). The inset shows the $\w^3$ dependence at small frequencies. Right: Spectral function at finite temperature below the BKT transition temperature.} 
\label{fig_chi_N=2} 
\end{figure} 

The spectral function for $N=2$ is shown in Fig.~\ref{fig_chi_N=2} for various values of $r_0$ and $T=0$. Again we observe a data collapse when $\Delta^{-\theta}\chi''(\w)$ is plotted as a function of $\w/\Delta$. In the low-energy limit, the flow equation~(\ref{rgeq}) is fully determined by the Goldstone mode, which leads to $\chi''(\w)\propto\w^3$ as observed in Fig.~\ref{fig_chi_N=2} (see inset). For $\w\gg\Delta$, we find the critical scaling $\chi''(\w)\sim \w^\theta$ with $\theta$ the exponent determined from the scaling of $\chi(i\wn)$ (Fig.~\ref{fig_chi_matsubara}). There is a well defined Higgs resonance whose location $\w=m_H$ and full width at half maximum vanish as the QCP is approached ($r_0\to r_{0c}$). The universal ratio $m_H/\Delta\simeq 2.4$ is compatible with the QMC estimates $2.1(3)$~\cite{Gazit13} and $3.3(8)$~\cite{Chen13}. Up to a multiplicative factor which depends on the nonuniversal prefactor 
$\calA_-$ (and was determined neither in the QMC simulations~\cite{Gazit13,Chen13} nor in the present approach), the shape of the resonance, given by the universal scaling function $\Phi_-$, is in very good agreement with the QMC result of Ref.~\cite{Gazit13}. This gives strong support to the validity of our NPRG approach. 

Figure~\ref{fig_chi_N=2} shows that the resonance persists at finite temperatures below the BKT transition temperature when $T\ll T_{\rm BKT}\simeq 0.42\Delta$. At frequencies $\w\gg T$, temperature has no noticeable effect: the behavior of the system is essentially quantum and the spectral function (including the Higgs resonance near $\w\sim m_H\gg T$) is well approximated by its $T=0$ value. At frequencies $\w\ll T$, the system behaves classically and the $\w^3$ dependence of the spectral function at $T=0$ is modified. In this frequency range, the numerical procedure to perform the analytic continuation becomes questionable. Nevertheless, noting that the spectral function is dominated by the Goldstone mode when $\w\ll\Delta$, we can use perturbation theory to obtain 
\beq
\chi''(\w) \sim \w^3 \coth\left(\frac{\w}{2T}\right) . 
\eeq
for $\w,T\ll \Delta$. We conclude that $\chi''(\w)\sim T\w^2$ at low frequencies $\w\ll T$. 

\begin{figure}
\centerline{\includegraphics[width=6cm]{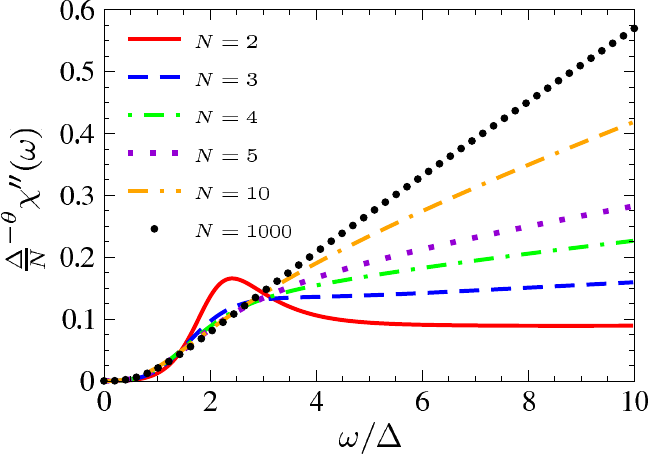}}
\caption{(Color online) $\Delta^{-\theta}\chi''(\w)/N$ vs $\w/\Delta$ for various values of $N$ in the $T=0$ ordered phase.} 
\label{fig_chi_N} 
\end{figure}

In Fig.~\ref{fig_chi_N} we show that the Higgs resonance is significantly suppressed for $N\geq 3$ (when we vary $N$ continuously between 2 and 3 we observe a gradual suppression). QMC simulations predict that the resonance is still marked for $N=3$~\cite{Gazit13,Gazit13a}. This discrepancy could be due to the limited precision of our method and more refined calculations are necessary to reach a definite conclusion regarding the precise form of $\Phi_-(x)$. For $N=1000$ we recover the exact $N\to\infty$ result~\cite{Podolsky12} showing no sign of a Higgs resonance.

\begin{figure}
\centerline{\includegraphics[width=6cm]{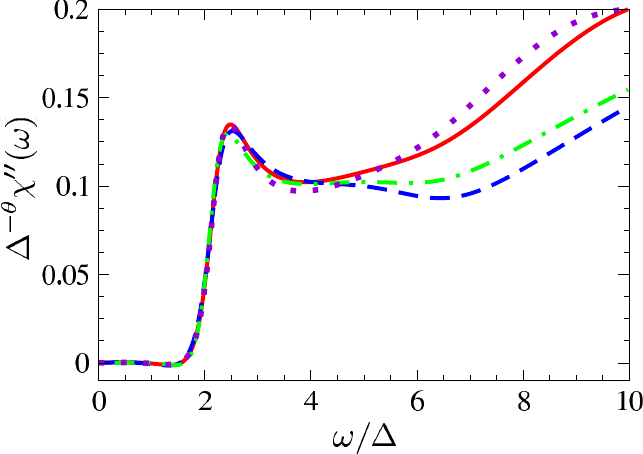}}
\caption{(Color online) $\Delta^{-\theta}\chi''(\w)$ vs $\w/\Delta$ in the $T=0$ disordered phase ($N=2$) for various values of $r_0$ ($0.001\leq r_{0c}-r_0\leq 0.005$ and $\Lamb=1$).} 
\label{fig_chi_QD_N=2and3} 
\end{figure} 

{\it Absence of Higgs resonance in the $T=0$ disordered phase.} Figure~\ref{fig_chi_QD_N=2and3} shows the spectral function $\chi''(\w)$ in the zero-temperature disordered phase for $N=2$. $\chi''(\w)$ vanishes for $\w<2\Delta$, rises sharply above the threshold, and exhibits a maximum for $\w\simeq 2.5\Delta$. Again these results are in good agreement with QMC simulations~\cite{Gazit13,Gazit13a,Pollet12,Chen13}. 
It has been suggested that the maximum observed in the spectral function can be interpreted as a Higgs resonance even though the system is disordered~\cite{Pollet12,Chen13,not7}. However, the behavior at length scales smaller than the 
correlation length $\xi$ is typical of a critical system~\cite{not6,not5}; at no length scales does the system behave as if it were ordered. This makes the $T=0$ disordered phase fundamentally different from the finite-temperature phase below the BKT transition even though both phases are characterized by the absence of true long-range order. 

\begin{figure}
\centerline{\includegraphics[height=4cm]{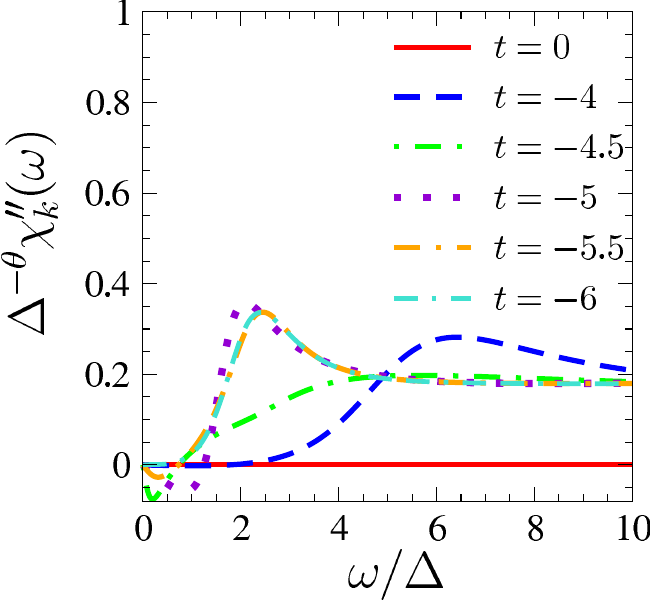}
\includegraphics[height=4cm]{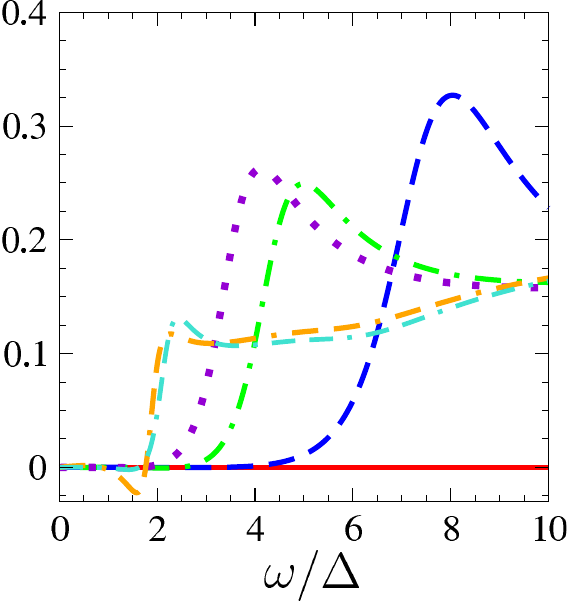}}
\caption{(Color online) $\chi_k''(\w)$ vs $\w/\Delta$ for various values of $t=\ln(k/\Lamb)$ ($T=0$ and $N=2$). Left panel: ordered phase ($t_J=
-\ln(\xi_J\Lambda)\simeq -5.2$). Right panel: disordered phase ($\rho_{0,k}$ vanishes for $t\simeq -3.9$).} 
\label{fig_chi_k} 
\end{figure}

To illustrate this point, let us discuss the spectral function $\chi_k''(\w)$ as a function of the RG momentum scale $k$. In the ordered phase, the RG flow exhibits a crossover between a critical regime $k\gg \xi_J^{-1}$ and a Goldstone regime $k\ll\xi_J^{-1}$, where $\xi_J=c/\rho_s\sim c/\Delta$ is the Josephson length~\cite{Rancon13a}. Figure~\ref{fig_chi_k} shows that the Higgs resonance is absent in the critical regime of the flow $k\gg\xi_J^{-1}$ and quickly builds up once $k$ reaches the Josephson scale $\xi_J^{-1}$. At nonzero but small temperature, $T\ll T_{\rm BKT}$, the flow exhibits a quantum-classical crossover for $k\sim T/c\ll\xi_J^{-1}$ which modifies the low-frequency behavior of the spectral function but not the Higgs resonance. 

In the disordered phase, the flow is critical as long as $\rho_{0,k}$ (the minimum of the effective action $\Gamma_k[\phibf]$) remains nonzero. At the value $k=k_c\sim \xi^{-1}$ for which $\rho_{0,k}$ vanishes, the maximum above the gap is still not formed. It builds up for $k<k_c$ when the RG flow is in the disordered phase ($\rho_{0,k}=0$). This definitively rules out the interpretation of the peak in the spectral function as a Higgs 
resonance. By the same argument we can also rule out the existence of a Higgs resonance in the finite-temperature disordered phase (quantum critical and quantum disordered regimes).

{\it Conclusion.} In summary, we have calculated the scalar susceptibility $\chi(\w)$ in the vicinity of a $(2+1)$-dimensional QCP. Besides the confirmation that a Higgs resonance is present for $N=2$ in the $T=0$ ordered phase, and the calculation of the universal properties of the spectral function $\chi''(\w)$ in agreement with quantum Monte Carlo simulations, we have shown that the resonance persists at finite temperatures below the BKT transition temperature. The spectral function then shows a $T\w^2$ dependence for $\w\ll T$.  We have also studied the possibility that a Higgs resonance exists in the zero- or finite-temperature disordered phase (characterized by a finite correlation length $\xi$): our RG analysis unambiguously reveals that the maximum observed in the spectral function cannot be interpreted as a Higgs resonance. Finally, we have shown that the resonance is strongly suppressed when $N\geq 3$. For $N=3$, our result disagrees with Monte Carlo simulations and more refined calculations are 
necessary to reach a definite conclusion.

\begin{acknowledgments}
We thank D. Podolsky and S. Gazit for useful correspondence. 
\end{acknowledgments}

%\bibliography{/users/lptl/dupuis/publi/BIB/nprg.bib,/users/lptl/dupuis/publi/BIB/bosons.bib,/users/lptl/dupuis/publi/BIB/book.bib,/users/lptl/dupuis/publi/BIB/notes.bib,/users/lptl/dupuis/publi/BIB/stat_phys.bib,/users/lptl/dupuis/publi/BIB/bcsbec.bib,/users/lptl/dupuis/publi/BIB/cold_atoms.bib,/users/lptl/dupuis/publi/BIB/fermions.bib,/users/lptl/dupuis/publi/BIB/supra.bib,/users/lptl/dupuis/publi/BIB/magnetism.bib}

%\bibliography{/home/nicolas/publi/BIB/nprg.bib,/home/nicolas/publi/BIB/bosons.bib,/home/nicolas/publi/BIB/book.bib,/home/nicolas/publi/BIB/notes.bib,/home/nicolas/publi/BIB/stat_phys.bib,/home/nicolas/publi/BIB/bcsbec.bib,/home/nicolas/publi/BIB/cold_atoms.bib,/home/nicolas/publi/BIB/fermions.bib,/home/nicolas/publi/BIB/supra.bib,/home/nicolas/publi/BIB/magnetism.bib}

\newpage

%AUTHOR'S MACRO
\def\Il{I_{k,\rm l}}
\def\It{I_{k,\rm t}}
\def\Jtt{J_{k,\rm tt}}
\def\Jll{J_{k,\rm ll}}
\def\Jtl{J_{k,\rm tl}}
\def\Jlt{J_{k,\rm lt}}
\def\bDA{\bar\Delta_{A,k}(p)}
\def\bDB{\bar\Delta_{B,k}(p)}

%%%%%%%%%%%%%%%%%%%%%%%%%%%%%%%%%%%%%%%%%%%

\section{Supplementary Material} 

\setcounter{equation}{0}
%\begin{abstract}
%\end{abstract}

We discuss the nonperturbative renormalization-group (NPRG) calculation of the scalar susceptibility in the two-dimensional relativistic quantum O($N$) model. In Sec.~\ref{sec_nprg} we briefly review the NPRG approach. We then describe a (simplified) Blaizot-Mendez-Galain (BMW) approximation (Sec.~\ref{sec_bmw}). Finally, in Sec.~\ref{sec_chi} we show how the scalar susceptibility $\chi(\w)$ can be computed in the NPRG approach.

\subsection{NPRG approach} 
\label{sec_nprg} 

The strategy of the NPRG approach is to build a family of theories indexed by a momentum scale $k$ such that fluctuations are smoothly taken into account as $k$ is lowered from the microscopic scale $\Lambda$ (the UV cutoff of the quantum O($N$) model) down to 0~\cite{Berges02a,Delamotte12a,Kopietz_booka}. This is achieved by adding to the action $S[\varphibf,h]$ of the quantum O($N$) model an external source term $-\int dx\, \J\cdot\varphibf$ ($x=(\r,\tau)$) and an infrared regulator term 
\beq
\Delta S_k[\varphibf] = \half \sum_{p,i} \varphi_i(-p) R_k(p) \varphi_i(p) , 
\eeq
where $i$ runs from 1 to $N$ and $p=(\p,i\wn)$ ($\wn$ denotes a bosonic Matsubara frequency). We choose an exponential cutoff function 
\begin{equation}
R_k(p) = Z_{A,k} \left(\p^2 + \frac{\wn^2}{c_k^2} \right) r \left( \frac{\p^2+\wn^2/c_k^2}{k^2} \right)
\end{equation}
with $r(x)=1/(e^x-1)$. $Z_{A,k}$ is the $k$-dependent field renormalization factor and $c_k$ denotes the renormalized ($k$-dependent) velocity of the $\varphibf$ field. The partition function of the system becomes $k$-dependent,
\beq
Z_k[\J,h] = \int \calD[\varphibf]\, e^{-S[\varphibf,h] - \Delta S_k[\varphibf]+\int dx \, \J\cdot\varphibf} ,
\eeq
while the order parameter is defined by 
\beq
\phibf_k[x;\J,h] = \frac{\delta\ln Z_k[\J,h]}{\delta\J(x)} . 
\label{phidef}
\eeq

The central quantity in the NPRG approach is the scale-dependent effective action
\beq
\Gamma_k[\phibf,h] = - \ln Z_k[\J,h] + \int dx\, \J\cdot \phibf - \Delta S_k[\phibf] ,
\label{gammadef} 
\eeq
defined as a modified Legendre transform of the free energy $-\ln Z_k[\J,h]$ which includes the subtraction of $\Delta S_k[\phibf]$. In Eq.~(\ref{gammadef}), $\J(x)\equiv\J_k[x;\phibf,h]$ is obtained by inverting Eq.~(\ref{phidef}). The variation of the effective action with $k$ is given by Wetterich's equation~\cite{Wetterich93a} 
\beq
\dt \Gamma_k[\phibf,h] = \half \Tr\llbrace \dt R_k\left(\Gamma^{(2)}_k[\phibf,h] + R_k\right)^{-1} \rrbrace ,
\label{rgeq2}
\eeq
where $\Gamma^{(2)}_k[\phibf,h]$ denotes the second-order functional derivative of $\Gamma_k[\phibf,h]$ with respect to $\phibf$ and $t=\ln(k/\Lambda)$. In Fourier space, the trace involves a sum over momenta and Matsubara frequencies as well as the O($N$) index of the $\phibf$ field. Since $R_{k=0}$ vanishes, the effective action $\Gamma[\phibf,h]\equiv \Gamma_{k=0}[\phibf,h]$ of the quantum O($N$) model is recovered for $k=0$. On the other hand, assuming that all fluctuations are frozen for $k=\Lambda$, $\Gamma_\Lambda[\phibf,h]=S[\phibf,h]$. 

\subsection{BMW approximation} 
\label{sec_bmw} 

To find an approximate solution of the flow equation~(\ref{rgeq2}) when $h=0$, we use the Blaizot-Mendez-Wschebor (BMW) approximation~\cite{Blaizot06a,Benitez09a,Benitez12a}. The latter is based on a RG equation for the 2-point vertex in a constant (i.e. uniform and time-independent) field, 
\beq
\Gamma^{(2)}_{k,ij}(p;\phibf) = \delta_{i,j} \Gamma_{A,k}(p;\rho) + \phi_i \phi_j \Gamma_{B,k}(p;\rho) , 
\eeq
where $\Gamma_{A,k}(p;\rho)$ and $\Gamma_{B,k}(p;\rho)$ are function of the O($N$) invariant $\rho=\phibf^2/2$. The RG equation of $\Gamma^{(2)}_{k,ij}$ involves the 3- and 4-point vertices: 
\begin{widetext}
\begin{multline}
\dt \Gamma_{k,ij}^{(2)}(p;\phibf) = - \half \sum_{q,i_1\cdots i_3} G_{k,i_1i_3}(q;\phibf) \dt R_k(q) G_{k,i_3i_2}(q;\phibf) \Gamma^{(4)}_{k,iji_2i_1}(p,-p,q,-q;\phibf) \\
+ \sum_{q,i_1\cdots i_5}  G_{k,i_1i_5}(q;\phibf) \dt R_k(q) G_{k,i_5i_2}(q;\phibf) \Gamma^{(3)}_{k,ii_2i_3}(p,q,-p-q;\phibf)  
G_{k,i_3i_4}(p+q;\phibf) \Gamma^{(3)}_{k,ji_4i_1}(-p,p+q,-q;\phibf) .
\label{rgeq1}
\end{multline}
\end{widetext}
Because of the $\dt R_k(q)$ term in Eq.~(\ref{rgeq1}), the integral over the loop variable $q$ is dominated by $|\q|\lesssim k$ and $|\wn|\lesssim c_k k$. Since the vertices are regular functions of their arguments when $|\p_i|\ll k$ and $|\w_{n_i}|\ll c_kk$ (a consequence of the infrared regulator term $\Delta S_k$ in the action), we can set $q=0$ in $\Gamma_k^{(3)}$ and $\Gamma_k^{(4)}$. Noting that in a constant field, 
\begin{equation}
\begin{split}
\Gamma_{k,i_1i_2i_3}^{(3)}(k;p,-p,0;\phibf) &= \frac{1}{\sqrt{\beta V}} \frac{\partial}{\partial\phi_{i_3}} \Gamma^{(2)}_{k,i_1i_2}(p,-p;\phibf) , \\
\Gamma_{k,i_1\cdots i_4}^{(4)}(p,-p,0,0;\phibf) &= \frac{1}{\beta V} \frac{\partial^2}{\partial\phi_{i_3}\partial\phi_{i_4}} \Gamma^{(2)}_{k,i_1i_2}(p,-p;\phibf) ,
\end{split}
\end{equation}
we then obtain a closed equation for $\Gamma_k^{(2)}(p,-p;\phibf)$~\cite{Blaizot06a,Benitez09a,Benitez12a}. 

It is convenient to introduce the self-energies $\Delta_{A,k}$ and $\Delta_{B,k}$ defined by 
\beq
\begin{split}
\Gamma_{A,k}(p;\rho) &= p^2 + U_k'(\rho) + \Delta_{A,k}(p;\rho) , \\ 
\Gamma_{B,k}(p;\rho) &= U_k''(\rho) + \Delta_{B,k}(p;\rho) ,
\end{split}
\eeq
where $\Delta_{A,k}(p=0;\rho)=\Delta_{B,k}(p=0;\rho)=0$ and $p^2=\p^2+\wn^2/c^2$. The effective potential $U_k$ is defined by 
\beq
U_k(\rho) = \frac{1}{\beta V} \Gamma_k[\phibf] \Bigl|_{\phibf=\const} 
\eeq
and is a function of $\rho$ because of the O($N$) symmetry. It satisfies the RG equation 
\beq
\dt U_k(\rho) = \half \int_q \dt R_k(q) \bigl[G_{k,\rm l}(q;\rho)+(N-1)G_{k,\rm t}(q;\rho)] ,
\label{Urgeq} 
\eeq
where 
\beq 
\begin{split}  
G_{k,\rm l}(p;\rho) &= [ \Gamma_{A,k}(p;\rho) + 2 \rho \Gamma_{B,k}(p;\rho) + R_k(p) ]^{-1} , \\ 
G_{k,\rm t}(p;\rho) &= [ \Gamma_{A,k}(p;\rho) + R_k(p) ]^{-1} ,
\end{split} 
\eeq
are the longitudinal and transverse parts of the propagator $G_k=(\Gamma_k^{(2)}+R_k)^{-1}$
in a constant field. We use the notation $\int_q=\frac{1}{\beta} \sum_{\wn} \int_\q$. 

The BMW approximation leads to functional RG equations for $p$- and $\rho$-dependent functions. In order to simplify the numerical solution, we consider two additional approximations. On the one hand we expand the effective potential about the position $\rho_{0,k}$ of its minimum,
\beq
U_k(\rho) = U_k(\rho_{0,k}) + \delta_k (\rho-\rho_{0,k}) + \frac{\lamb_k}{2} (\rho-\rho_{0,k})^2 .
\eeq
On the other hand we approximate the self-energies by their values $\bar\Delta_{A,k}(p)$ and $\bar\Delta_{B,k}(p)$ at $\rho_{0,k}$, 
\beq
\begin{split}
\bar\Delta_{A,k}(p) &\equiv \Gamma_{A,k}(p;\rho_{0,k}) - p^2 - \Gamma_{A,k}(0;\rho_{0,k}) , \\ 
\bar\Delta_{B,k}(p) &\equiv \Gamma_{B,k}(p;\rho_{0,k}) - \Gamma_{B,k}(0;\rho_{0,k}) . \\ 
\end{split}
\eeq
This leads to the flow equations 
\begin{align}
\dt \bDA =  - 2 \rho_{0,k} \lamb_k^2 [ & \Jtl(p) + \Jlt(p) \nonumber \\ &
 - \Jtl(0) - \Jlt(0) ] , 
\label{rgeqA}
\end{align}
and
\begin{align}
\dt \bDB ={}& -(N-1)[\lamb_k+\bDB]^2 \Jtt(p) \nonumber \\ 
& + \lamb_k^2 [ \Jlt(p) + \Jtl(p) - 9 \Jll(p) - \Jlt(0)  \nonumber \\ 
& - \Jtl(0) + 9\Jll(0) + (N-1) \Jtt(0) ] ,
\label{rgeqB}
\end{align}
where the threshold function $I_{k,\alpha}$ and $J_{k,\alpha\beta}$ ($\alpha,\beta={\rm l,t}$) are defined by 
\begin{equation}
\begin{split} 
I_{k,\alpha} &= \int_q \tdt G_{k,\alpha} (q;\rho_{0,k}) , \\ 
J_{k,\alpha\beta}(p) &= \int_q [\tdt G_{k,\alpha}(q;\rho_{0,k})] G_{k,\beta}(p+q;\rho_{0,k}) .
\end{split}
\end{equation}
The operator $\tdt=(\dt R_k)\partial_{R_k}$ acts only on the $k$ dependence of the cutoff function $R_k$. To a good approximation, we can use a derivative expansion (DE) of the propagator in the threshold functions, 
\beq
\begin{split}
\bar\Delta^{\rm DE}_{A,k}(p) &= (Z_{A,k}-1)\p^2 + \left(V_{A,k}-\frac{1}{c^2} \right) \wn^2 , \\ 
\bar\Delta^{\rm DE}_{B,k}(p) &= 0 . 
\end{split}
\eeq 
This approximation has been shown to be reliable for the classical O($N$) model~\cite{Benitez08a,Sinner08a}. 

\subsection{The scalar susceptibility} 
\label{sec_chi} 

The scalar susceptibility is defined by 
\beq
\chi(p) = \frac{\delta^2 \ln Z[h]}{\delta h(-p) \delta h(p)} \biggl|_{h=0}
= - \frac{\delta^2 \Gamma[\bar\phibf[h],h]}{\delta h(-p) \delta h(p)} \biggl|_{h=0} , 
\label{chi2} 
\eeq
where the order parameter $\bar\phibf[h]$ is defined by 
\beq
\frac{\delta\Gamma[\phibf,h]}{\delta\phibf(x)} = 0. 
\eeq
In Eq.~(\ref{chi2}), $\delta/\delta h(p)$ is a total derivative which acts both on $\bar\phibf[h]$ and the explicit $h$-dependence of the functional $\Gamma[\phi,h]$. We deduce 
\beq
\chi(p) = - \bar\Gamma^{(0,2)}(p) + \bigl[\bar\Gamma_{\rm l}^{(1,1)}(p) \bigr]^2 \bar G_{\rm l}(p) ,
\eeq
where $\bar G_{\rm l}(p)=G_{\rm l}(p;\rho_{0})$ and we have introduced the vertices
\beq
\begin{split}
\bar\Gamma^{(0,2)}(p) &=  \frac{\delta^2 \Gamma[\phibf,h]}{\delta h(-p) \delta h(p)} \biggl|_{\phibf=\bar\phibf,h=0} , \\ 
\bar\Gamma_{\rm l}^{(1,1)}(p) &= \frac{\delta^2 \Gamma[\phibf,h]}{\delta\phi_1(-p) \delta h(p)} \biggl|_{\phibf=\bar\phibf,h=0} 
\end{split}
\eeq
with $\bar\phibf=(\sqrt{2\rho_{0}},0\cdots 0)$ the order parameter for $h=0$. 

\subsubsection{Truncation of the effective action}

We use the following truncation of the effective action,
\begin{multline}
\Gamma_k[\phibf,h] = \Gamma_k[\phibf,0] +  \half \int dy dy'\, h(y) H_k^{(0,2)}(y-y') h(y') \\
+ \int dx dy [\rho(x)-\rho_{0,k}] H^{(2,1)}_k(x-y) h(y) .
\label{ansatz2}
\end{multline} 
The $k$-dependent scalar susceptibility is then expressed as 
\beq
\chi_k(p) = -  H_k^{(0,2)}(p) + 2 \rho_{0,k} [ H^{(2,1)}_k(p) ]^2 \bar G_{k,\rm l}(p) ,
\eeq 
while $H_k^{(0,2)}$ and $H^{(2,1)}_k$ satisfy the RG equations 
\beq
\begin{split}
\dt H_k^{(0,2)}(p) ={}& - [ \Jll(p) + (N-1) \Jtt(p) ] [ H^{(2,1)}_k(p) ]^2 ,  \\ 
\dt  H_k^{(2,1)}(p) ={}& - \lbrace \Jll(p)[3\lamb_k+2\bDB] 
\\ & + (N-1) \Jtt(p) [\lamb_k+\bDB] \rbrace H^{(2,1)}_k(p) 
\end{split} 
\label{rgeqH}
\eeq
in the ordered phase. Similar equations can be obtained in the disordered phase.

\subsubsection{Large-$N$ limit and Goldstone regime} 

In the large-$N$ limit, only transverse fluctuations contribute to the flow equations~(\ref{rgeqA},\ref{rgeqB},\ref{rgeqH}). Integrating the latter in the superfluid phase, one finds 
\begin{align}
\chi_k(p) ={}& 12 \frac{N}{u_0} 
- \left( \frac{6N}{u_0}\right)^2 \nonumber \\ &  \times \frac{p^2}{p^2\left( \frac{3N}{u_0} + \frac{N}{2} [ \Pi_k(p) - \Pi_\Lambda(p)] \right) + 2\rho_0 } , 
\end{align} 
where
\beq
\Pi_k(p) = \int_q \frac{1}{[q^2+R_k(q)][(p+q)^2+R_k(p+q)]} .
\eeq
Neglecting $\Pi_\Lambda(p)$ with respect to $\Pi_{k=0}(p)=c/[8(\p^2+\wn^2/c^2)^{1/2}]$, we recover the known result $\chi_{k=0}(p)$ of the large-$N$ limit. The corresponding spectral function  does not show any Higgs resonance in the scaling limit but varies as $\w^3$ in the low frequency limit. 

In the superfluid phase, the infrared limit of the flow when $k\ll \xi_J^{-1}\sim \Delta/c$ ($\xi_J$ denotes the Josephson length) is dominated by the $N-1$ Goldstone modes and longitudinal fluctuations can be ignored. The flow equations are then similar to those of the large-$N$ limit (with however $N-1$ instead of $N$). This ensures that the spectral function $\chi''(\w)$ obtained from the NPRG equations vanishes as $\w^3$ for $\w\to 0$. 

\subsubsection{Critical scaling}

The exponent $\theta$ of the spectral function $\chi''(\w)\sim\w^\theta$ in the critical regime differs from $3-2/\nu$. There are two sources of error in the exponent $\theta$. On the one hand our result for the correlation-length exponent $\nu$ differs from the exact value: $\nu\simeq 0.613$ instead of $\nu\simeq 0.6717$ for $N=2$, and $\nu\simeq 0.699$ instead of $\nu\simeq 0.7112$ for $N=3$. On the other hand, our approximations also lead to an error on the scaling dimension of $H_k^{(2,1)}$. The equality $\theta=3-2/\nu$ is obtained if $H^{(2,1)}$ has scaling dimension $[H^{(2,1)}]=2-\eta-1/\nu$ where $\eta$ the anomalous dimension at the quantum critical point. It is not satisfied in our approach. We find $[H^{(2,1)}]\simeq 0.517$ instead of $[H^{(2,1)}]\simeq 0.31$ (using $\nu\simeq 0.613$ and $\eta\simeq 0.0582$ with our approximations) for $N=2$ and $[H^{(2,1)}]\simeq 0.62$ instead of $[H^{(2,1)}]\simeq 0.52$ (using $\nu\simeq 0.699$ and $\eta\simeq 0.0507$) for $N=3$.

%\bibliography{/home/nicolas/publi/BIB/nprg.bib,/home/nicolas/publi/BIB/bosons.bib,/home/nicolas/publi/BIB/book.bib,/home/nicolas/publi/BIB/notes.bib,/home/nicolas/publi/BIB/stat_phys.bib,/home/nicolas/publi/BIB/bcsbec.bib,/home/nicolas/publi/BIB/cold_atoms.bib,/home/nicolas/publi/BIB/fermions.bib,/home/nicolas/publi/BIB/supra.bib,/home/nicolas/publi/BIB/magnetism.bib}

\begin{thebibliography}{34}
\bibitem [{not({\natexlab{a}})}]{not4}%
  \BibitemOpen
  \href@noop {} {} \bibinfo {note} {The universal regime is
  defined by $|\p|,|\w|/c\ll k_G$ where $k_G^{-1}\sim 24\pi/(cu_0)$ is the
  Ginzburg length.}\BibitemShut {Stop}%
\bibitem{not7} Note that this suggestion is not supported by the authors of Ref.~\cite{Gazit13,Gazit13a}.
\bibitem [{not({\natexlab{e}})}]{not6}%
  \BibitemOpen
  \href@noop {} {} \bibinfo {note} {The system exhibits a
  critical behavior at length scales $\xi_G\ll L \ll \xi$ where
  $\xi_G=k_G^{-1}$ is the Ginzburg length~\cite{not4}.}\BibitemShut {Stop}%
\bibitem{not5} No resonance can form when the system is critical, in agreement with the $\w^{3-2/\nu}$ behavior of the spectral function.
\end{thebibliography}


\end{document}

\end{document}